\documentclass{PoS}

\title{Review of recent highlights in lattice calculations at finite temperature and finite density}

\ShortTitle{Recent highlights in lattice calculations at T>0}

\author{\speaker{P. Petreczky}\thanks{This work was supported by U.S. Department of Energy under
Contract No. DE-AC02-98CH10886.}\\
        Physics Department, Brookhaven National Laboratory, 
         Upton, NY 11973, USA\\
        E-mail: \email{petreczk@bnl.gov}}

\abstract{

I review some recent lattice results on studying chiral and deconfinement transition in 
QCD at finite temperature and density, as well as properties of strongly interacting matter
at high temperatures. I discuss lattice results on the equation of state, fluctuations of 
conserved charges, chromo-electric screening as well as the determination of the 
chiral transition temperature.

          }

\FullConference{Xth Quark Confinement and the Hadron Spectrum,\\
		October 8-12, 2012\\
		TUM Campus Garching, Munich, Germany}

\begin{document}

\section{Introduction}
It is expected that strongly interacting matter undergoes a transition in some temperature
interval from hadron gas to deconfined state also called the quark gluon plasma (QGP) \cite{gros81}. 
Creating and exploring deconfined medium in the laboratory is 
the goal of the large experimental heavy ion program at RHIC 
\cite{nagle} and  LHC \cite{salgado}.
Studying hot and dense strongly interacting matter is also the subject of a large effort in lattice QCD 
(see Refs. \cite{myrev,owerev} for recent reviews). Early lattice QCD simulations at non-zero temperature  
were limited to large quark masses and had no control over the discretization errors 
\cite{milc_old_eos,wilson_old,p4_old}. During the past seven years calculations
with the physical strange quark mass and physical or nearly physical light ($u,d$) quark masses 
have been performed using improved staggered fermion actions 
\cite{fodor05,our_Tc,fodor06,milc_eos,our_eos,eos005,hotQCD,fodor09,fodor10,fodor10eos,hotqcd2}, 
and for several quantities continuum extrapolated results have been obtained.
There was also progress in lattice QCD calculations at non-zero temperature using other
fermion formulations, namely Wilson fermions \cite{wilson_thermo,whotqcd,mainz},
Domain-Wall fermions \cite{prasad} and overlap fermions \cite{overlap_thermo}. 
The later two formulations preserve the chiral symmetry of continuum QCD.
However, due to much larger computational costs of these formulations 
the corresponding results are far less extensive.

To get reliable predictions from lattice QCD the lattice spacing $a$ should be sufficiently small
relative to the typical QCD scale, i.e. $\Lambda_{QCD} a \ll 1$. For staggered fermions, discretization
errors go like ${\cal O}( (a \Lambda_{QCD})^2)$ but discretization errors due to flavor symmetry
breaking turn out to be quite large numerically, and dominate the cutoff dependence of thermodynamic
quantities at low temperatures.
To reduce these errors one has to use improved staggered fermion actions with so-called fat links
\cite{orginos}. At high temperature the dominant discretization errors come from the lattice 
distortions of the quark dispersion relation and go like $(a T)^2$,  and therefore
could be very large. Thus,  it is important to use improved discretization schemes, 
which reduce or eliminate these discretization errors. Staggered fermion actions 
used in numerical calculations typically implement some version of fat links as well as improvement
of quark dispersion relation and are referred to as $p4$, $asqtad$, $HISQ$ and $stout$. Independently
of specific improvement all lattice results eventually must be extrapolated to the continuum limit.

In this contribution I am going to discuss lattice QCD calculations on the equation of state, 
study of deconfinement aspects of the QCD transition, 
including color screening and fluctuations of conserved charges and determination of the chiral 
transition temperature. I will mostly discuss lattice results obtained with staggered quark formulation; 
where appropriate results obtained using other actions will be highlighted.

\section{Equation of State}
\begin{figure}[ht]
\includegraphics[width=7.3cm]{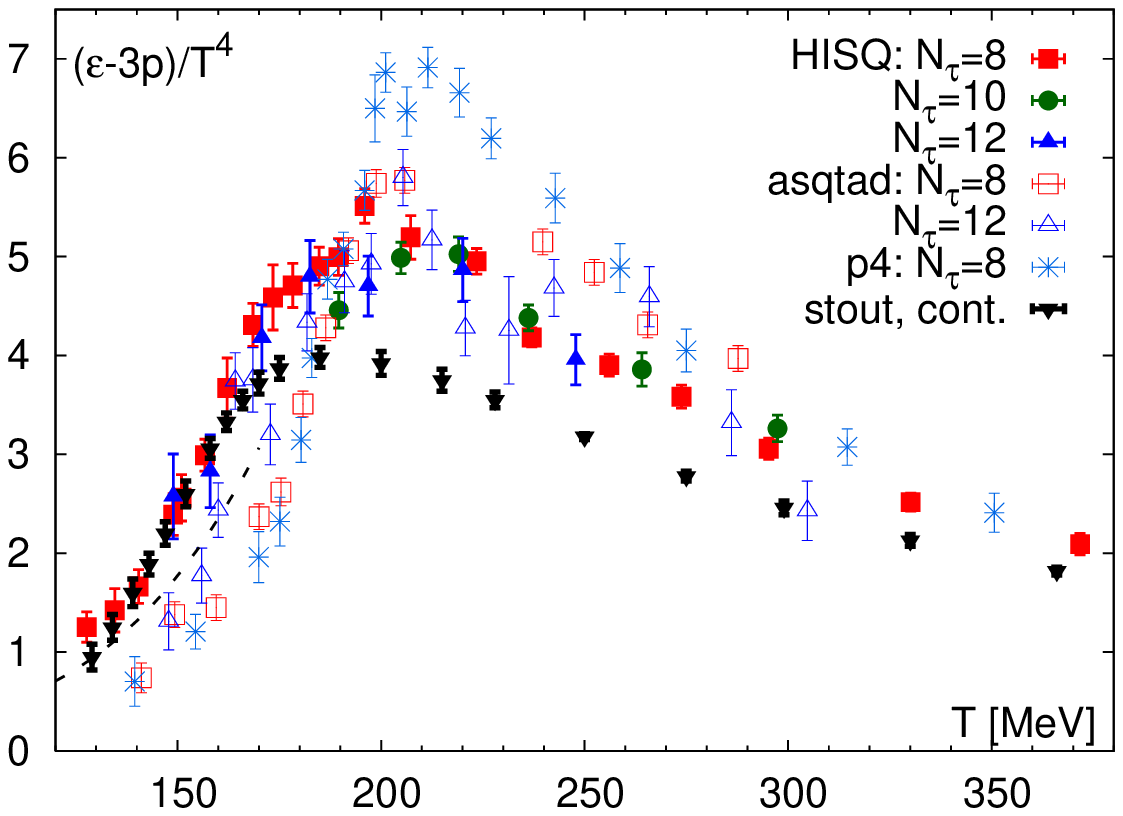}
\includegraphics[width=7.5cm]{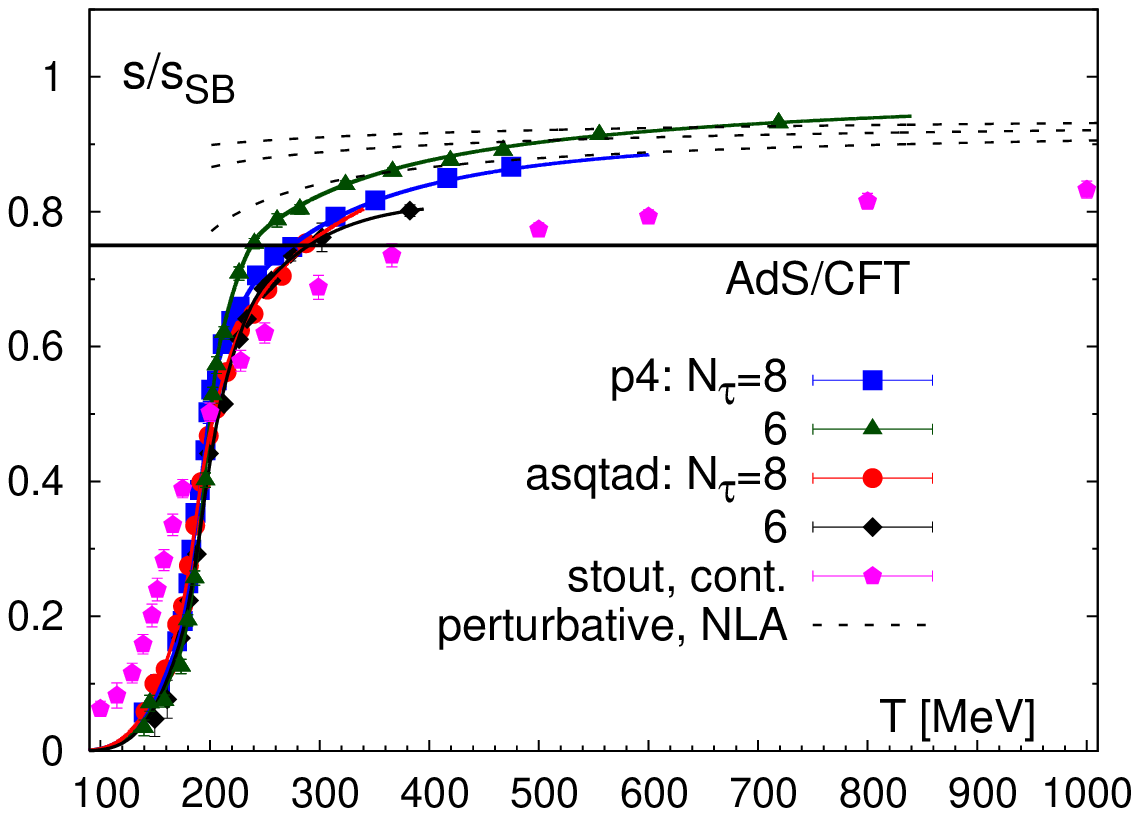}
\vspace*{-0.3cm}
\caption[]{
The interaction measure  (left) and the
entropy density (right) as function of the temperature calculated with improved staggered fermion actions.
The dashed line in the left panel shows the HRG result. The band in the right panel shows the resummed
perturbative result, while the solid line is the expectations based on the strongly coupled limit.
}
\label{fig:eos}
\end{figure}

The equation of state has been calculated using different improved staggered fermion actions
$p4$, $asqtad$, $stout$ and $HISQ$. 
In the lattice calculations of the equation of state and many other quantities
the temperature is varied by varying the lattice spacing at fixed value of the temporal extent $N_{\tau}$.
The temperature $T$ is given by the lattice spacing and the temporal extent, $T=1/(N_{\tau} a)$. Therefore
taking the continuum limit corresponds to $N_{\tau} \rightarrow \infty$ at the fixed physical volume.
The calculation
of thermodynamic observables proceeds through the calculation of the trace of the energy momentum tensor,
$\epsilon -3 p$, also
known as the trace anomaly or the interaction measure. This is due to the fact that this quantity can be expressed in
terms of expectation values of local gluonic and fermionic operators, (see e.g. Ref. \cite{hotQCD}).  
Different thermodynamic observables can be obtained through integration of
the trace anomaly
\footnote{A somewhat different approach was used in Ref. \cite{fodor10eos}}. 
The pressure can be written as
\begin{equation}
\displaystyle
\frac{p(T)}{T^4}-\frac{p(T_0)}{T_0^4}=\int_{T_0}^T \frac{dT'}{T'^5} (\epsilon-3 p).
\end{equation}
The lower integration limit $T_0$ is chosen such that the pressure is exponentially small there.
Furthermore, the entropy density can be written as $s=(\epsilon+p)/T$. Since the interaction measure 
is the basic thermodynamic observable in the lattice calculations it is worth to discuss its properties
more in detail. In Fig. \ref{fig:eos} (left panel) I show 
the results of the calculation with different improved actions. The calculation with $p4$ and $asqtad$
actions have been performed on $N_{\tau}=8$ lattices and light quark masses $m_l=m_s/10$, with $m_s$
being the physical strange quark mass \cite{our_eos,hotQCD}.
These light quark masses correspond to the pion masses slightly above $200$ MeV in the continuum
limit. For this value of $N_{\tau}$ the above deviation from the physical quark mass plays little role 
\cite{eos005,latproc}. The $N_{\tau}=12$ asqtad calculations have been performed for $m_l=m_s/20$ \cite{latproc12}.
Calculations with $HISQ$ action have been performed for $N_{\tau}=4,~6,~8,~10$ and
$12$ for $m_l=m_s/20$ corresponding to the pion mass of $160$ MeV in the continuum limit \cite{latproc12}.
A subset of these results is shown in Fig. \ref{fig:eos}. Finally, calculation
of the trace anomaly and the equation of state was performed with $stout$ action using $N_{\tau}=4,~6,~8,~10$ and
$12$ and physical light quark masses \cite{fodor10eos}. Using the lattice data from $N_{\tau}=6,~8$ and $10$ 
a continuum estimate for different quantities was given \cite{fodor10eos}. 
The interaction measure shows a rapid rise in the transition region and after reaching a peak at temperatures
of about $200$ MeV decreases. At low temperatures lattice data obtained with $HISQ$ action and stout action
agree with each other. We also compare the lattice results with hadron resonance gas (HRG) model calculations
which seems to agree well with $HISQ$ and $stout$ results for $T<150$ MeV.  
Cutoff effects (i.e. $N_{\tau}$ dependence) 
appears to be the strongest around the peak region.
They decrease at high temperatures and at $T>400$ MeV all lattice results agree with each other.
At low temperature the cutoff effects related to flavor symmetry breaking are very large for $p4$ and $asqtad$ actions.
Due to these large cutoff effects the $N_{\tau}=8$ $p4$ and $asqtad$ data are below the hadron resonance gas model. Taking into
account the distortions of the hadron spectrum due to flavor symmetry breaking in hadron resonance gas
calculations leads to good agreement of HRG model with the lattice \cite{pasi}. 
Since at high temperatures the effects of flavor symmetry breaking in the pressure and 
the interaction measure are small, the
reduction in $\epsilon-3p$ at low temperatures must be compensated (at least partly) by a larger value
at intermediate temperatures.  
Thus, the large $N_{\tau}$ dependence of the peak height of $\epsilon-3p$ is related to the large flavor symmetry breaking
effects for $p4$ and $asqtad$ actions.

In Fig. \ref{fig:eos} I also show the entropy density divided by the corresponding ideal gas value
and compare the lattice results with the resummed perturbative calculations \cite{blaizot,blaizot1}, 
as well as with the predictions from AdS/CFT correspondence for the strongly coupled regime \cite{gubser98}.
The later is considerably below the lattice results. Note that the pressure, the energy density 
and the trace anomaly have also been recently discussed in the framework of 
resummed perturbative calculations which seem to agree with lattice data quite well at
high temperatures\cite{mike}.
The differences between the $stout$ action and the $p4$ and $asqtad$ actions for the trace anomaly
translates into the differences in the pressure and the  entropy density. 
In the high temperature region the entropy density obtained with stout action is $10\%$ smaller
than the entropy density obtained with $p4$ and $asqtad$ actions.

\section{Taylor expansion of the pressure and  fluctuations of conserved charges}
Due to the infamous sign problem lattice QCD Monte-Carlo simulations are not possible at non-zero
quark chemical  potentials.
The pressure and other quantities at non-zero chemical potentials, however, can be evaluated using Taylor expansion. 
 The 
Taylor expansion can be set up in terms of the quark chemical potentials $\mu_u$, $\mu_d$ and
$\mu_s$,  or in terms of the chemical
potentials corresponding to baryon number $B$, electric charge $Q$ and strangeness $S$ of hadrons 
\begin{eqnarray}
\frac{p}{T^4}&=&\frac{1}{VT^3}\ln Z(T,\mu_u,\mu_d,\mu_s)=\sum_{ijk} \frac{1}{i! j! k!} \chi_{ijk}^{uds} 
\left(\frac{\mu_u}{T}\right)^i \left(\frac{\mu_d}{T}\right)^j \left(\frac{\mu_s}{T}\right)^k\nonumber\\
\chi_{ijk}^{uds}&=&\frac{\partial^{\,i+j+k}p/T^4}{\partial(\mu_{u}/T)^{i}\partial(\mu_{d}/T)^{j}\partial(\mu_{s}/T)^{k}}\\
\frac{p}{T^4}&=&\frac{1}{VT^3}\ln Z(T,\mu_B,\mu_Q,\mu_S)=\sum_{ijk} \frac{1}{i! j! k!} \chi_{ijk}^{BQS} 
\left(\frac{\mu_B}{T}\right)^i \left(\frac{\mu_Q}{T}\right)^j \left(\frac{\mu_S}{T}\right)^k \nonumber\\
\chi_{ijk}^{BQS}&=&\frac{\partial^{\,i+j+k}p/T^4}{\partial(\mu_{B}/T)^{i}\partial(\mu_{Q}/T)^{j}\partial(\mu_{S}/T)^{k}}.
\end{eqnarray}
Using Taylor expansion method the equation of state has been calculated for small chemical potential 
in the continuum limit \cite{fodor_eosmub}. Earlier results at non-zero lattice spacing
have been reported in Refs. \cite{ejiri_eosmub,milc_eos_mub}.
While Taylor expansion can be used to study the physics at non-zero baryon density, the
expansion coefficients are interesting on 
their own right as they are related to fluctuations and correlations of conserved charges.
As will become clear later fluctuations and correlations of conserved charges are good probes of deconfinement.
Fluctuations of conserved charges are also useful for determining the freeze-out conditions in heavy
ion experiments \cite{freezout}.

\begin{figure}
\hspace*{-0.3cm}
\includegraphics[width=5.2cm]{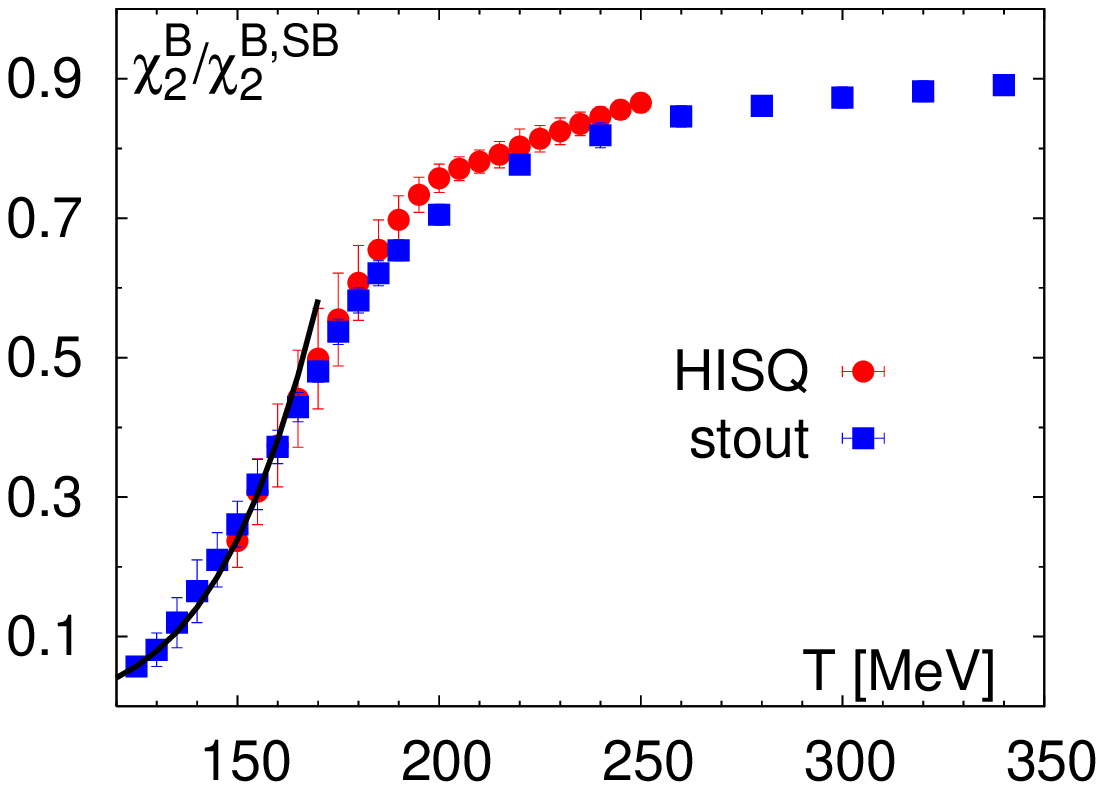}
\hspace*{-0.3cm}
\includegraphics[width=5.2cm]{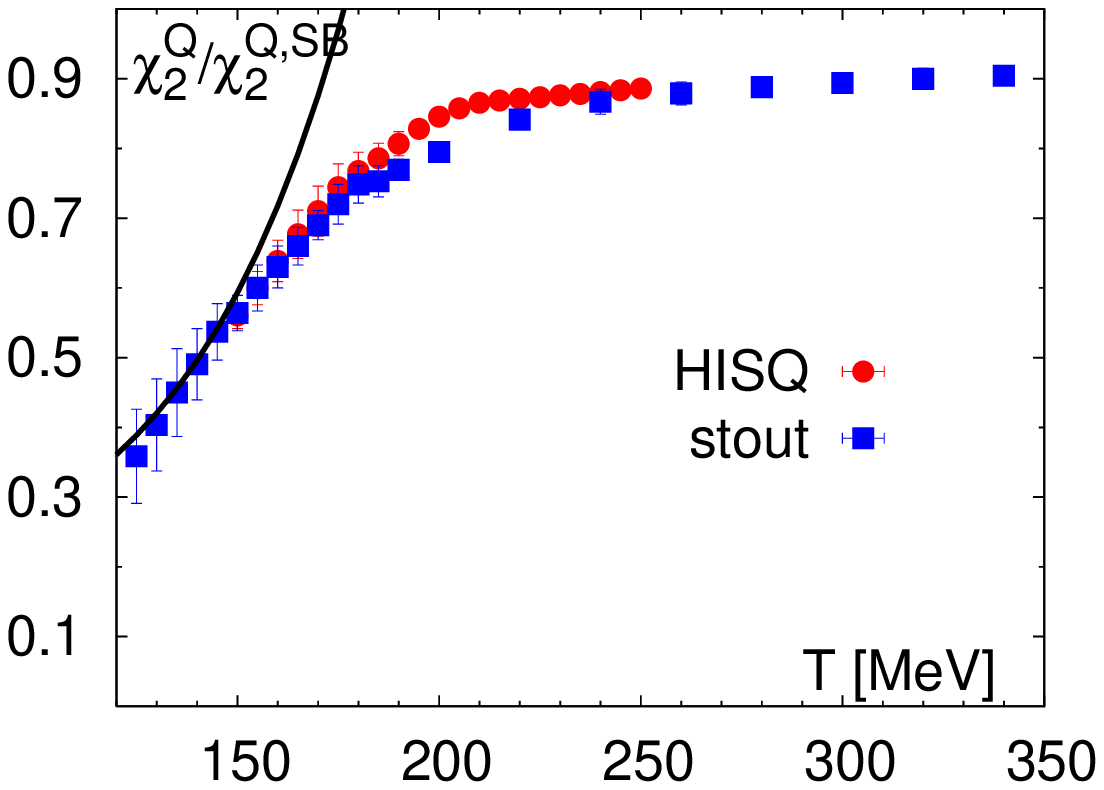}
\hspace*{-0.3cm}
\includegraphics[width=5.2cm]{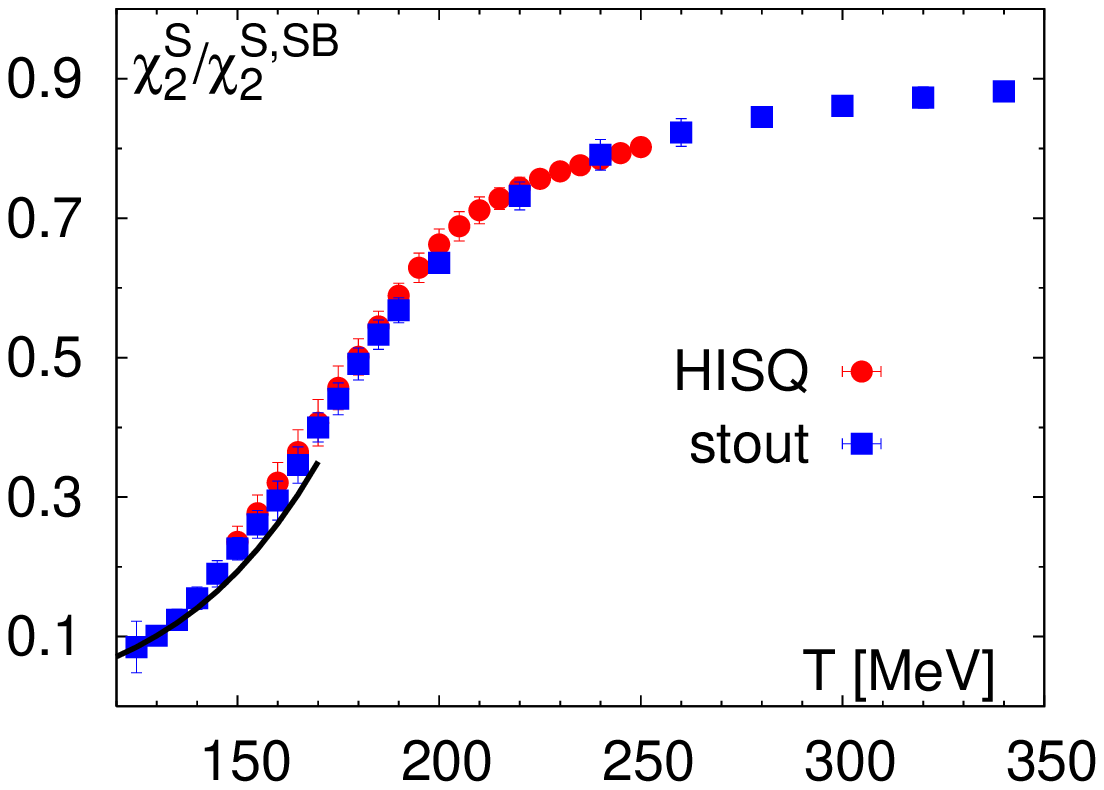}
\caption{The fluctuations of baryon number (left), electric charge (middle) and strangeness (right)
as function of the temperature calculated with $HISQ$ and stout action in the continuum limit
and normalized by the corresponding ideal quark gas values $\chi_2^{X,SB}$. The solid black
curves correspond to HRG predictions.}
\label{fig:chi2comp}
\end{figure}

\begin{figure}
\includegraphics[width=7.7cm]{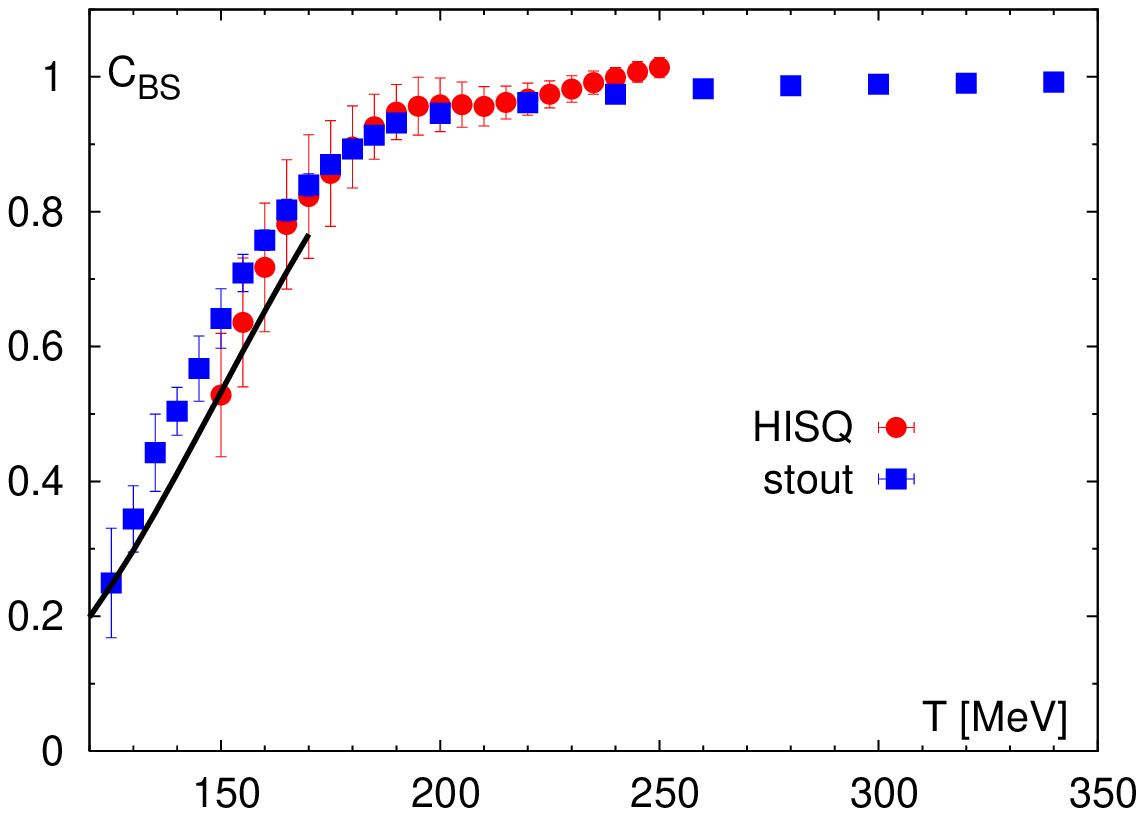}
\includegraphics[width=7.7cm]{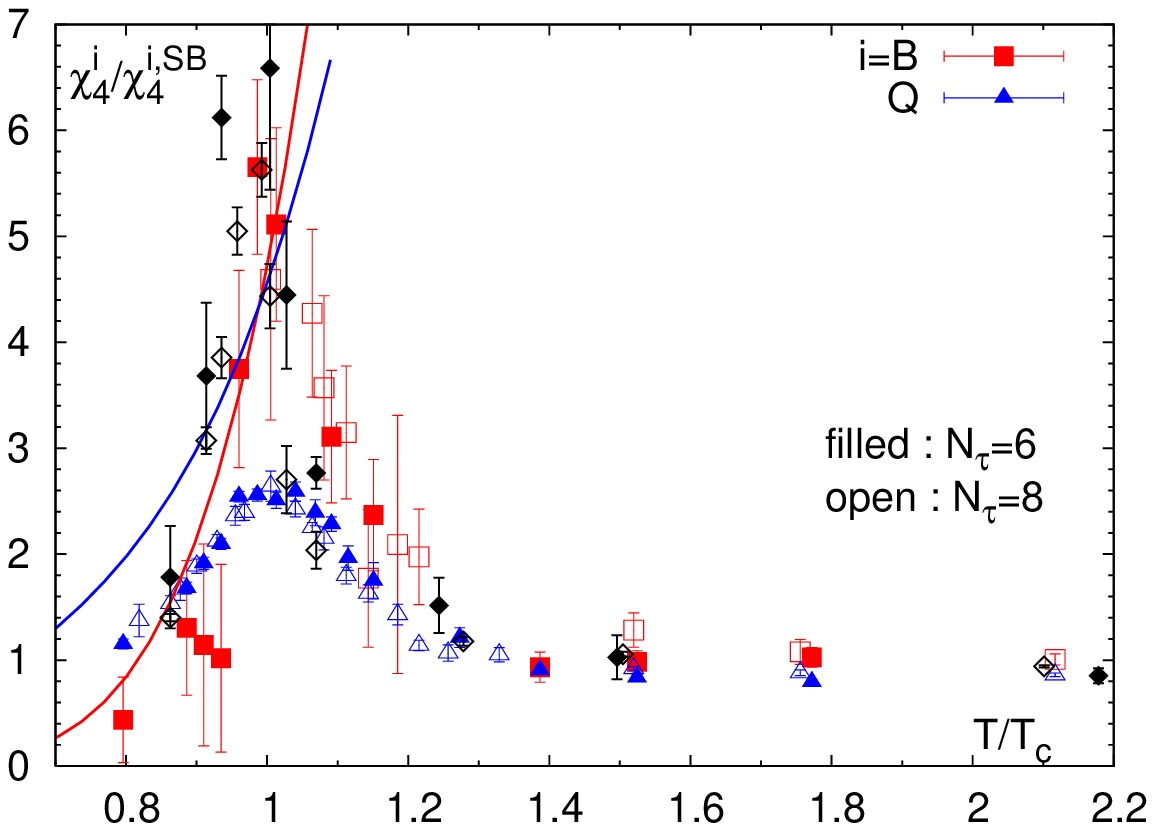}
\caption{Baryon number-strangeness correlation $C_{BS}$ (left) and fourth order fluctuations
of baryon number and electric charge for $HISQ$ action normalized by the corresponding ideal quark gas value 
as function of $T/T_c$ (right). The lines correspond to the prediction of the hadron resonance gas model.
Open (filled) diamonds in the right panel correspond to $N_{\tau}=4$ $p4$ results for $\chi_4^Q$ ($\chi_4^B$).
}
\label{fig:chi4}
\end{figure}

The diagonal expansion coefficients are related to second and higher order fluctuations of conserved charges, e.g.
\begin{eqnarray}
\chi_2^X &=& \frac{1}{VT^3}\langle N_X^2\rangle \nonumber \\
\chi_4^X &=& \frac{1}{VT^3}\left(\langle N_X^4\rangle - 
3 \langle N_X^2\rangle^2\right) \;\; ,
\label{fluc}
\end{eqnarray}
while the off-diagonal expansion coefficients are related to correlations among conserved charges, e.g.
\begin{equation}
\chi_{11}^{XY}=\frac{1}{VT^3}\langle N_XN_Y\rangle.
\end{equation}
Second order fluctuations have been studied with improved staggered actions in Refs. \cite{hotQCD,fodor10,our_fluct}.
Recently continuum results have been obtained for second order fluctuations of baryon number, electric charge
and strangeness using $stout$ and $HISQ$ actions \cite{fodor_fluct,hotqcd_fluct} which are shown in Fig. \ref{fig:chi2comp}.
The lattice results are also compared with HRG model.
As one can see from the figure at low temperatures the lattice data are described well 
by HRG model indicating that the dominant degrees of freedom in that temperature  range are hadronic.
Deconfiniment is seen as a rapid increase of the fluctuations for $T>150$ MeV, which eventually reach values
that are close to the expectations of weakly interacting quark gas. In other words, the fluctuations indicate quark
degrees of freedom at high temperatures. The lattice results obtained with $HISQ$ and s$tout$ action agree well in the continuum
limit, except for temperatures around $200$ MeV where some discrepancies are observed in the baryon and electric charge
fluctuations. Another way to study deconfinement is to consider correlations of
conserved charges. These correlations are very different for hadron gas and quark gas. As an example let us
examine the strangeness-baryon number correlation. It is convenient to normalize this quantity as follows
$C_{BS}=-3 \chi_{11}^{BS}/\chi_S$. At high temperatures where strangeness is carried by s-quarks this quantity
should be close to one. At low temperatures on the other hand strange baryons are responsible for strangeness-baryon correlations.
In Fig. \ref{fig:chi4}(left) I show continuum results for $C_{BS}$ obtained with $stout$ \cite{fodor_fluct} and
$HISQ$ \cite{hotqcd_fluct} actions. At low temperatures the lattice results are described by HRG model, while
at high temperatures  they are close to one as expected for  quark gas.

Fourth order fluctuations for baryon number, electric charges and strangeness have been studied
using $p4$ \cite{our_fluct} and $HISQ$ \cite{recent_proc} actions. Cutoff effects are quite large for $p4$ action in the low temperature
and the transition regions. These large cutoff effects also result in much larger value of the transition temperature
than obtained with $stout$ and $HISQ$ actions \cite{our_Tc}. Calculations of the fourth and higher order fluctuations 
are quite demanding computationally and for this reason no continuum results have been obtained yet.
The lattice results obtained with $HISQ$ action for $N_{\tau}=6$ and $8$ are shown in Fig. \ref{fig:chi4} 
as function of the temperature in units of the chiral transition temperature $T_c=154$ MeV (see below) 
and also compared with
earlier results obtained with $p4$ action and $N_{\tau}=4$. For the $p4$ action we use the value of
the transition temperature $T_c=204$ MeV determined in Ref. \cite{our_Tc} for $N_{\tau}=4$.
To reduce the cutoff effects the lattice spacing in the $HISQ$ calculations
was set by the kaon decay constant $f_K$. At high temperatures the fourth order fluctuations are close to the values corresponding
to non-interacting quark gas. At low temperatures the fourth order baryon number fluctuations are reasonably well described by 
hadron resonance gas. This is not the case, however, for the electric charge. One possible reason for the disagreement between
the lattice and HRG model in this case could be the large cutoff effects in the pion sector. 
Electric charge fluctuations are very sensitive to
the pion sector, which is largely distorted on the lattice even if $HISQ$ or $stout$ action is used. These distortions correspond
effectively to  a larger pion mass that would explain why the lattice data are below the HRG expectations.
Fourth order fluctuations have a maximum in the transition region. Interestingly enough the position of the maximum
in $T/T_c$ is roughly the same for the $HISQ$ and $p4$ actions. The height of the peak, however, is significantly larger
for the $\chi_4^Q$ when $p4$ action is used.

Second and fourth order light and strange quark number fluctuations at high temperatures have been studied on
the lattice in Refs. \cite{fodor_fluct,my_lat09}
and compared with resummed perturbative calculations \cite{vuorinen,mustafa}. The resummed perturbative calculations
seem to describe the lattice data quite well at $T>300$ MeV.

\section{Deconfinement : color screening}
\begin{figure}
\includegraphics[width=8cm]{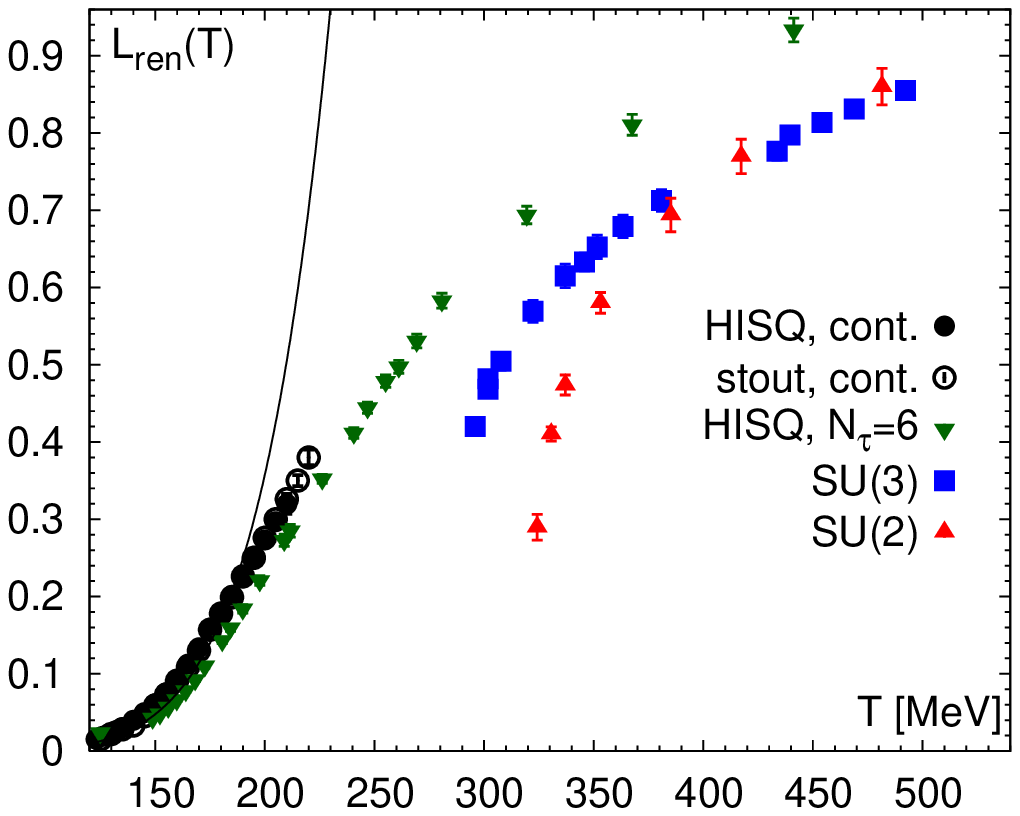}
\includegraphics[width=7.5cm]{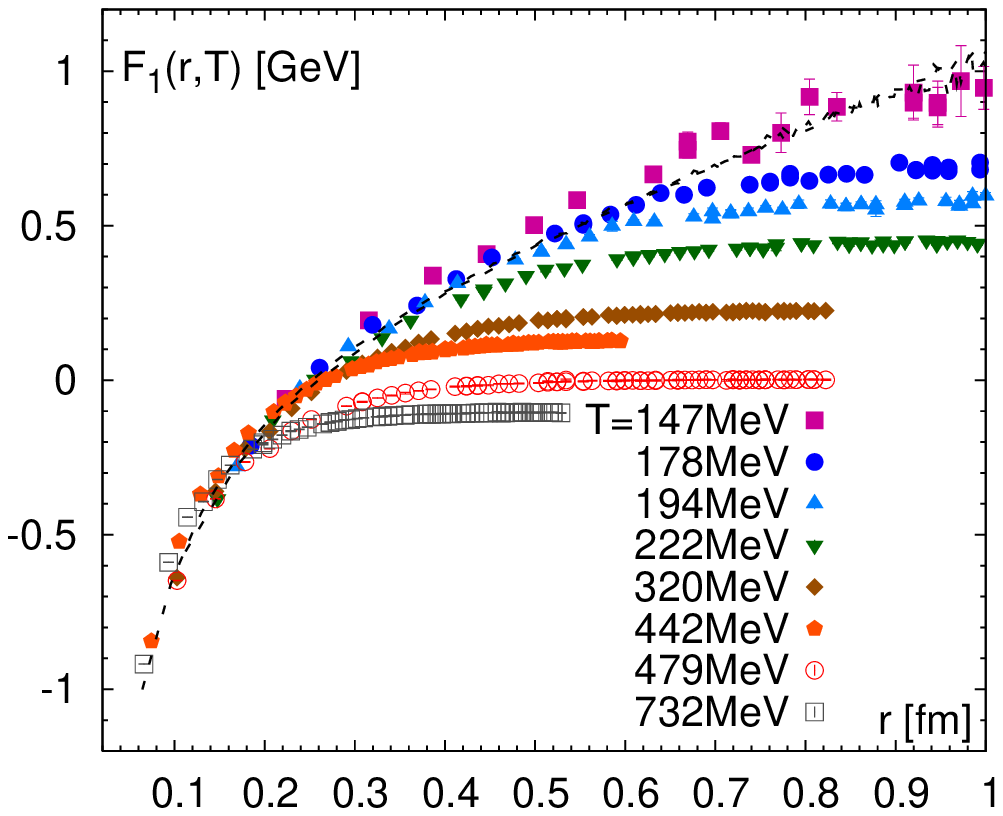}
\caption{Left: The renormalized Polyakov loop as function of the temperature in 2+1 flavor QCD
and pure gauge theory. Right: The singlet free energy as function of the distance at
different temperature calculated with $HISQ$ action \cite{f1_new}.
The solid line in the left panel corresponds to static-ligh(strange) hadron gas
prediction for $L_{ren}$ (see text). The dashed line in the right panel is the $T=0$ potential \cite{hotqcd2}. }  
\label{fig:f1} 
\end{figure}

One of the most prominent feature of the quark gluon plasma is the presence of chromoelectric (Debye) screening.
The easiest way to study chromoelectric screening is to calculate the Polyakov loop.
The Polyakov loop is an order parameter for the deconfinement transition in pure gauge theory,
which is governed by $Z(N)$ symmetry. For QCD this symmetry is explicitly broken
by dynamical quarks. There is no obvious reason for the Polyakov loop
to be sensitive to the singular behavior close to the chiral limit, although speculations
along these lines have been made \cite{speculations}. The Polyakov loop 
is related to the screening properties of the medium and thus to deconfinement.
After proper renormalization, the square of the Polyakov loop characterizes the
long distance behavior of the static quark anti-quark free energy; it 
gives the excess in free energy needed to screen two well-separated color
charges. 
The renormalized Polyakov loop, calculated on lattices 
with temporal extent $N_\tau$, is obtained from the bare Polyakov 
\begin{eqnarray}
&
\displaystyle
L_{ren}(T)=z(\beta)^{N_{\tau}} L_{bare}(\beta)=
z(\beta)^{N_{\tau}} \left<\frac{1}{3}  {\rm Tr } W(\vec{x}) \right >,~~W(\vec{x})=
\prod_{x_0=0}^{N_{\tau}-1} U_0(x_0,\vec{x}),
\end{eqnarray}
where $U_0=\exp(iga A_0)$ denotes the temporal gauge link and $z(\beta)$ 
is the renormalization constant determined from the $T=0$ static potential \cite{fodor06}.
Continuum results for the renormalized Polyakov loop have been obtained 
with $stout$ \cite{fodor10} and $HISQ$ actions \cite{lren_new}.
These are shown in  Fig.~\ref{fig:f1} together with $N_{\tau}=6$ $HISQ$ results \cite{hotqcd2}.
One can see a good agreement
between the $stout$ and $HISQ$ results. 
I also compare
the 2+1 flavor QCD results with the corresponding results in pure gauge theory \cite{okacz02,digal03}
as well as with the prediction of non-interacting gas of static-light(strange) hadrons \cite{lren_new,megias}.
We see that in the vicinity of the transition temperature the behavior of the renormalized
Polyakov loop in QCD and in the pure gauge theory is quite different. The calculation of $L_{ren}$ based
on non-interacting static-light hadron gas can explain the lattice data for $T<140$ MeV.
The renormalized Polyakov loop has also been calculated using lattice fermion
formulations other than staggered, namely the Wilson formulation \cite{wilson_thermo} and the overlap formulation \cite{overlap_thermo}.
These formulations are considerably more expensive computationally than the staggered formulation and therefore
the calculations have been performed at unphysical pion mass. The results obtained  using Wilson action
and overlap action for the Polyakov loop agree very well with the staggered fermion results at the same value
of the pion masses \cite{wilson_thermo,overlap_thermo}.

Further insight on chromoelectric screening can be gained by studying
the singlet free energy of static quark anti-quark pair (for 
reviews on this see Ref. \cite{mehard04,qgp09}),  which is expressed in terms of 
the correlation function of temporal Wilson lines in Coulomb gauge
\begin{equation}
\exp(-F_1(r,T)/T)=\frac{1}{3} {\rm Tr} \langle W(r) W^{\dagger}(0) \rangle.
\end{equation}
Instead of using the Coulomb gauge the singlet free energy can be defined in gauge invariant manner by
inserting a spatial gauge connection between the two Wilson lines. Using such definition the singlet free energy 
has been calculated in $SU(2)$ gauge theory \cite{baza08}.  
It has been found that the singlet free energy calculated this way is close to the result obtained in
Coulomb gauge \cite{baza08}. 
The singlet free energy turned out to be  useful to 
study quarkonia binding at high temperatures in potential models 
(see e.g. Ref. \cite{mocsy1} and references therein).  It
also appears naturally in the perturbative
calculations of the Polyakov loop correlators at short distances \cite{plc_new}.

The  singlet free energy has been recently calculated 
in 2+1 flavor QCD with $HISQ$ action on $24^3 \times 6$ and $16^3 \times 4$ lattices \cite{f1_new}. 
The numerical results are shown in Fig. \ref{fig:f1}.
At short distances the singlet free energy is temperature independent and coincides with the 
zero temperature
potential.  In purely gluonic theory the free energy grows linearly with the separation between 
the static quark and 
anti-quark in the confined phase. In presence of dynamical quarks the free energy is saturated 
at some finite value 
at distances of about $1$ fm due to string breaking (see e.g. Ref. \cite{mehard04}). 
This is also seen in Fig. \ref{fig:f1}. 
Above the deconfinement temperature the singlet free
energy is exponentially screened at sufficiently large 
distances \cite{okacz02,digal03} with the screening mass proportional to
the temperature , i.e.
\begin{equation}
F_1(r,T)=F_{\infty}(T)-\frac{4}{3}\frac{g^2(T)}{4 \pi r} \exp(-m_D(T) r), ~m_D \sim T.
\end{equation}
The lattice data for the singlet free energy are consistent wit these expectations for
$r>0.8/T$.

Let me finally note that contrary to the electro magnetic plasma the static chormomagnetic fields
are screened in QGP. This is due to the fact that unlike photons gluons interact with each other
(the stress tensor is non-linear in QCD). Magnetic screening is non-perturbative, i.e. it does not
appear at any finite order of pertubation theory. In lattice calculations chromomagnetic screening 
is studied either in terms of spatial Wilson loops \cite{sigma_s} or in terms of spatial gluon
propagators \cite{prop98,prop00,prop01}. The numerical results obtained so far show that the length
scale related to magnetic screening is larger than the one related to electric screening.

\section{Chrial transition}
The Lagrangian of QCD  has an approximate $SU_A(3)$ chiral symmetry. This symmetry is broken in the 
vacuum. The chiral symmetry breaking is signaled by  non-zero expectation value of the quark or chiral condensate,
$\langle \bar \psi \psi \rangle \neq 0$ in the massless limit.
This symmetry is expected to be restored at high temperatures and the quark condensate vanishes.
There is an explicit breaking of the chiral symmetry by the non-zero values of $u,d$ and $s$ quark masses. 
While due to the relatively large strange quark mass ($m_s \simeq 100$ MeV) $SU_A(3)$ may not
be a very good symmetry its subgroup $SU_A(2)$ remains a very good symmetry and is relevant
for the discussion of the finite temperature transition in QCD. If the relevant symmetry is 
$SU_A(2)$ the chiral transition is expected to be second order for massless light ($u$ and $d$) quarks
belonging to the $O(4)$ universality class. Recent calculations with $p4$ action support this pictures \cite{MEoS}.
This also means that for non-zero light quark masses the transition must be a crossover. The crossover nature
of the transition is supported by calculations in Ref. \cite{nature}.
The $U_A(1)$ symmetry is explicitly
broken in the vacuum by the anomaly but it is expected to be effectively restored at high temperatures
as non-perturbative vacuum fluctuations responsible for its breaking are suppressed at high temperatures.
If the  $U_A(1)$ symmetry is restored at the same temperature as the  $SU_A(2)$ symmetry the transition
could be first order \cite{pisarski81}. Recent calculations with staggered \cite{mscr} as well as with domain wall fermions \cite{prasad}
suggest that $U_A(1)$  symmetry gets effectively restored at temperature that is significantly higher than the chiral
transition temperature.

For massless quark the chiral condensate vanishes at the critical temperature $T_c^0$ and is the order
parameter. Therefore in the lattice studies one calculates the chiral condensate and its derivative
with respect to the quark mass called the chiral susceptibility. For the staggered fermion formulation
most commonly used in the lattice calculations at non-zero temperature these quantities can be written as follows:
\begin{eqnarray}
\langle \bar \psi \psi \rangle_{q,x}&=&\frac{1}{4} \frac{1}{N_{\sigma}^3 N_{\tau}} 
{\rm Tr} \langle D_q^{-1} \rangle ,\\
\chi_{m,q}(T)&=& 
n_f \frac{\partial \langle \bar\psi \psi \rangle_{q,\tau}}{\partial m_l}
=\chi_{q, disc} + \chi_{q, con} ~~q=l,s, \label{susc}
\end{eqnarray}
where the subscript $x=\tau$ and $x=0$ will denote the expectation value
at finite and zero temperature, respectively.  
Furthermore, $D_q=m_q \cdot 1 + D$ is the fermion matrix in the canonical normalization and
$n_f=2$ and $1$ for light and strange quark.
In Eq. (\ref{susc}) we made explicit that chiral susceptibility is the sum of connected 
and disconnected Feynman diagrams. The disconnected and connected contributions can be written
as
\begin{eqnarray}
\chi_{q, disc} &=&
{{n_f^2} \over 16 N_{\sigma}^3 N_{\tau}} \left\{
\langle\bigl( {\rm Tr} D_q^{-1}\bigr)^2  \rangle -
\langle {\rm Tr} D_q^{-1}\rangle^2 \right\}
\label{chi_dis} \; , \\
\chi_{q, con} &=&  -
{{n_f} \over 4} {\rm Tr} \sum_x \langle \,D_q^{-1}(x,0) D_q^{-1}(0,x) \,\rangle \; ,~~~q=l,s.
\label{chi_con}
\end{eqnarray}
The disconnected part of the light quark susceptibility describes the
fluctuations in the light quark condensate and is directly analogous to
the fluctuations in the order parameter of an $O(N)$ spin model. The
second term ($\chi_{q,con}$) arises from the explicit quark mass
dependence of the chiral condensate and is the expectation value of
the volume integral of the correlation function of the (isovector)
scalar operator $\bar{\psi}\psi$. Let me note that in the massless limit only
$\chi_{l,disc}$ diverges. 

\subsection{The temperature dependence of the chiral condensate}
\begin{figure}
\includegraphics[width=7cm]{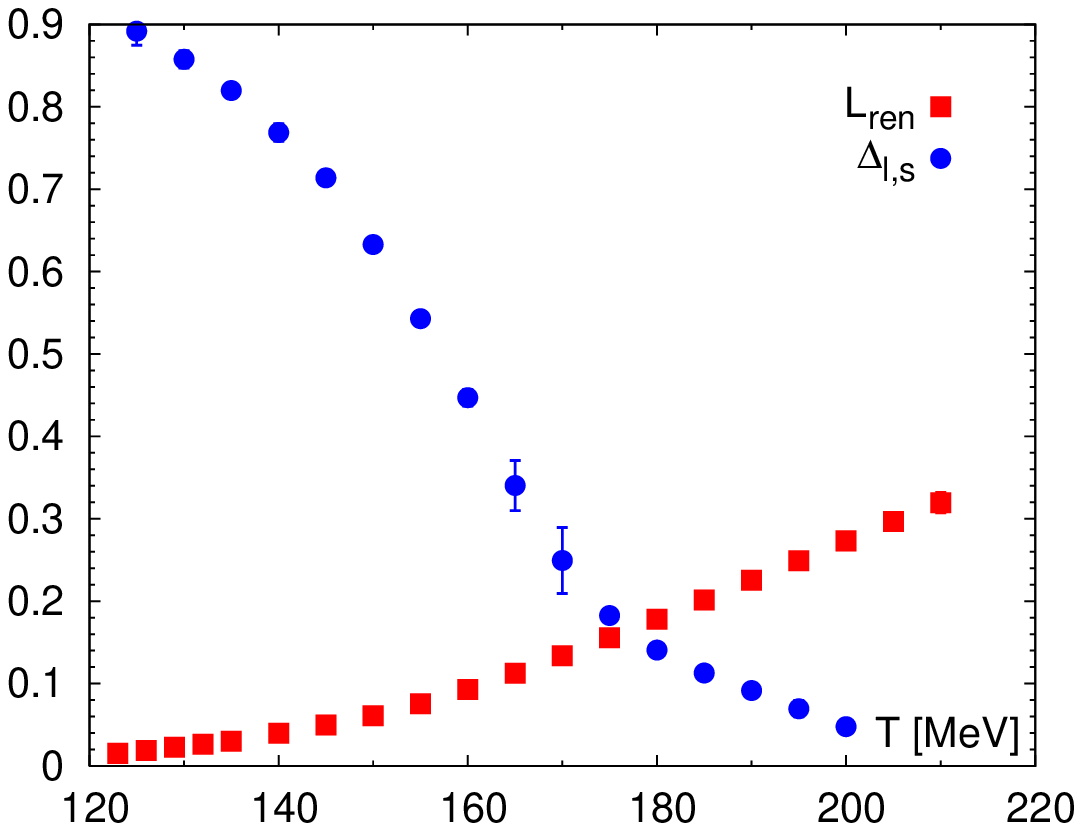}
\includegraphics[width=7cm]{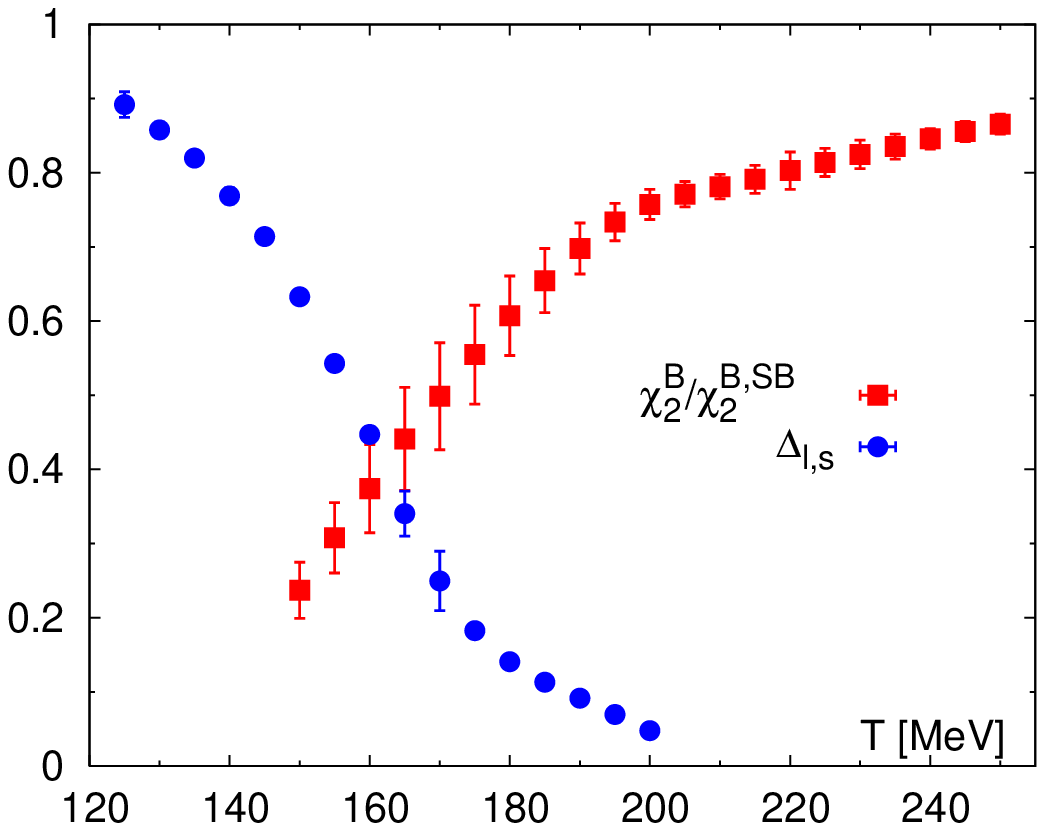}
\caption{The subtracted chiral condensate calculated with $HISQ$ action in the continuum limit 
compared to the renormalized Polyakov loop (left) and light quark number fluctuation (right).
The continuum results for $\Delta_{l,s}$ and $L_{ren}$ have been taken from Ref. \cite{lren_new},
while the continuum results for $\chi_2^B$ have been taken from Ref. \cite{hotqcd_fluct}.}
\label{fig:Delta}
\end{figure}

The chiral condensate needs a multiplicative, and also an additive renormalization if
the quark mass is non-zero. Therefore the subtracted chiral condensate is 
considered
\begin{equation}
\Delta_{l,s}(T)=\frac{\langle \bar\psi \psi \rangle_{l,\tau}-\frac{m_l}{m_s} \langle \bar \psi \psi \rangle_{s,\tau}}
{\langle \bar \psi \psi \rangle_{l,0}-\frac{m_l}{m_s} \langle \bar \psi \psi \rangle_{s,0}}.
\end{equation}
In Fig. \ref{fig:Delta} I show continuum results for $\Delta_{l,s}$ calculated with $HISQ$ action and compared to
the renormalized Polyakov loop and baryon number fluctuation previously discussed in relation to the deconfining
transition. 
The rapid increase in $\chi_2^B$ happens roughly in the same temperature interval where $\Delta_{l,s}$ shows
a rapid decrease, while it is difficult to make similar statements for $L_{ren}$ due to its very smooth
behavior. However, it is clear from Fig. \ref{fig:Delta} that $L_{ren}$ is very far from unity for temperatures
where $\Delta_{l,s}$ is very small.

\begin{figure}
\includegraphics[width=0.450\textwidth]{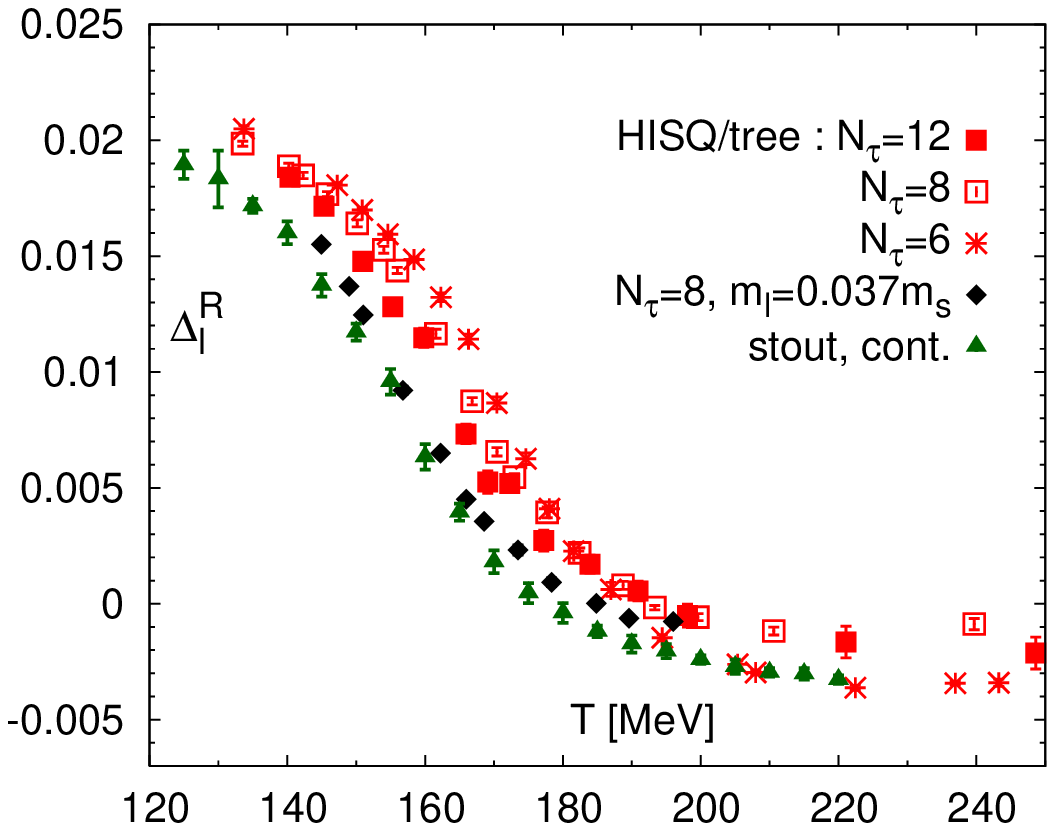}
\includegraphics[width=0.450\textwidth]{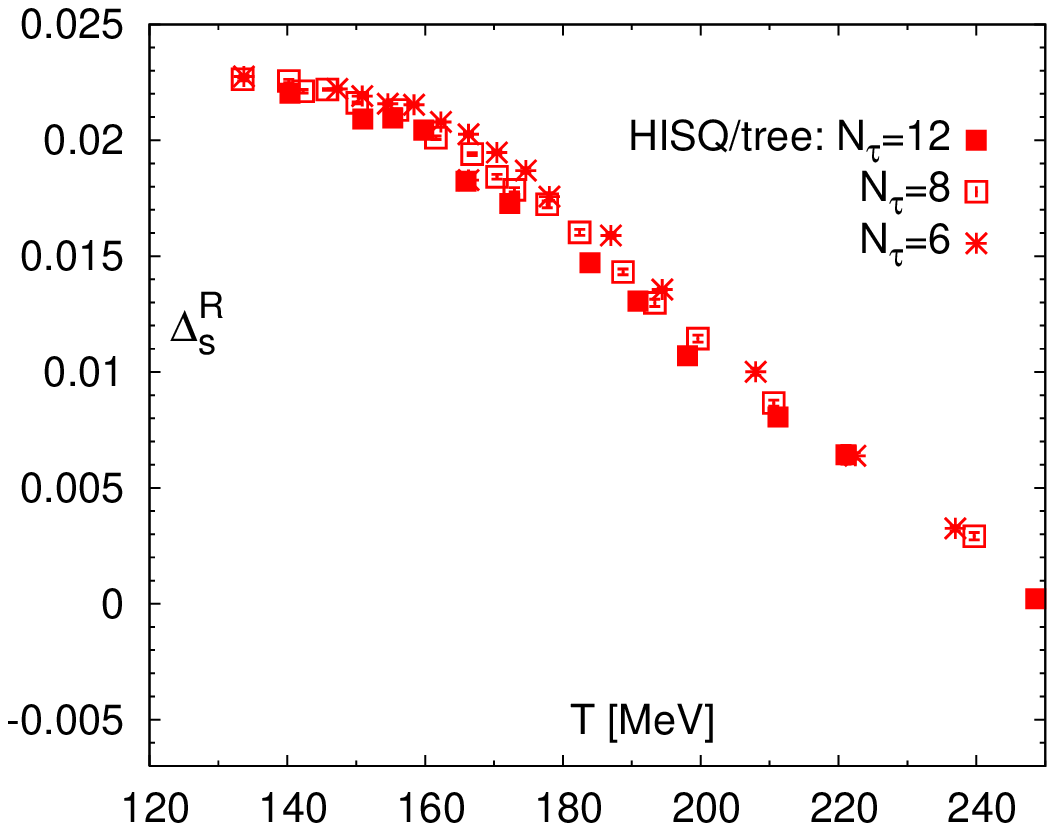}
\caption{The renormalized chiral condensate $\Delta_l^R$ for the
  $HISQ$ action with $m_l/m_s = 0.05$ is compared to the $stout$ data.
  In the right panel, we show the renormalized strange quark
  condensate $\Delta_s^R$ for the $HISQ$ action.
}
\label{fig:pbpR}
\end{figure}

Another way to get rid of the multiplicative and additive renormalization is to subtract the zero
temperature condensate and multiply the difference by the strange quark mass, i.e. consider the following
quantity
\begin{equation}
\Delta_q^R=d+2 m_s r_1^4 ( \langle \bar \psi \psi \rangle_{q,\tau}-
\langle \bar \psi \psi \rangle_{q,0} ),~~~ q=l,s .
\end{equation}
The factor $r_1^4$ was introduce to make the combination dimensionless. Here $r_1$ is the
scale parameter defined from the zero temperature static potential \cite{hotqcd2}.
It is convenient to choose the normalization constant to be the light quark condensate for $m_l=0$
multiplied by $m_s r_1^4$.  In Fig. \ref{fig:pbpR} the renormalized quark condensate is shown as function of
the temperature for $HISQ$ and $stout$ actions. We see a crossover behavior for temperature of $(150-160)$ MeV,
where $\Delta_l^R$ drops by $50\%$. The difference between the $stout$ and $HISQ$ results is a quark mass
effect. Calculations for $HISQ$ action were performed for $m_{\pi}=160$ MeV, while the $stout$ calculations
were done for the physical quark mass. For a direct comparison with $stout$ results, we extrapolate the 
$HISQ$ data in the light quark mass and also take care of the residual cutoff dependence in the $HISQ$
data. This was done in Ref. \cite{hotqcd2} and the results are shown in the figure as black diamonds demonstrating
a good agreement between $HISQ$ and $stout$ results.
Contrary to $\Delta_l^R$ the renormalized strange quark
condensate $\Delta_s^R$ shows only a gradual decrease over a wide temperature interval dropping by $50\%$
only at significantly higher temperatures of about $190$ MeV.
The subtracted chiral condensate has also been calculated using Wilson and 
overlap formulations \cite{wilson_thermo,overlap_thermo}. These calculations show good agreement
with staggered results at the corresponding values of the pion mass.

\subsection{O(N) scaling and the transition temperature}

In the vicinity of the chiral phase transition, the free energy
density may be expressed as a sum of a singular and a regular
parts,
\begin{equation}
f = -\frac{T}{V} \ln Z\equiv f_{sing}(t,h)+ f_{reg}(T,m_l,m_s) \; .
\label{free_energy}
\end{equation}
Here $t$ and $h$ are dimensionless couplings that control deviations from
criticality. They are related to the temperature $T$ and the light quark mass $m_l$ as  
\begin{equation}
t = \frac{1}{t_0}\frac{T-T_c^0}{T_c^0} \quad , \quad 
h= \frac{1}{h_0} H \quad , \quad 
H= \frac{m_l}{m_s} \; ,
\label{reduced}
\end{equation}
where $T_c^0$ denotes the chiral phase transition temperature, 
{\it i.e.}, the transition temperature at $H=0$.  The scaling variables
$t$, $h$ are normalized by two parameters $t_0$ and $h_0$, which are
unique to QCD and similar to the low energy constants in the chiral
Lagrangian.  These need to be determined together with $T_c^0$. In the
continuum limit, all three parameters are uniquely defined, but depend
on the value of the strange quark mass.

The singular contribution to the free energy density is a homogeneous
function of the two variables $t$ and $h$. Its invariance under scale
transformations can be used to express it in terms of a single
scaling variable
\begin{equation}
z=t/h^{1/\beta\delta} = \frac{1}{t_0}\frac{T-T_c^0}{T_c^0} \left( \frac{h_0}{H} \right)^{1/\beta\delta}
 = \frac{1}{z_0}\frac{T-T_c^0}{T_c^0} \left( \frac{1}{H} \right)^{1/\beta\delta}
\label{eq:defz}
\end{equation}
where $\beta$ and $\delta$ are the critical exponents of the $O(N)$
universality class and $z_0 = t_0/h_0^{1/\beta\delta}$.
Thus, the dimensionless free energy density
$\tilde{f}\equiv f/T^4$ can be written as
\begin{equation}
\tilde{f}(T,m_l,m_s) = h^{1+1/\delta} f_f(z) + f_{reg}(T,H,m_s) \; ,
\label{scaling}
\end{equation}
where $f_f$ is the universal scaling function and the regular term $f_{reg}$ 
gives rise to scaling violations. This regular term 
can be expanded in a Taylor series around $(t,h)=(0,0)$.

It should be noted that the reduced temperature $t$ may depend on other
couplings in the QCD Lagrangian which do not explicitly break 
chiral symmetry. In particular, it depends on light and strange 
quark chemical potentials $\mu_q$, which in leading order enter only quadratically,
\begin{equation}
t = \frac{1}{t_0} \left( \frac{T-T_c^0}{T_c^0} + 
\sum_{q=l,s}\kappa_q\left(\frac{\mu_q}{T}\right)^2 +
\kappa_{ls} \frac{\mu_l}{T}\frac{\mu_s}{T} \right)
  \; .
\label{reduced2}
\end{equation}

The transition temperature can be defined as peaks in susceptibilities (response functions) that are 
second derivatives of the free energy density with respect to relevant parameters. Since there are
two relevant parameters we can define three susceptibilities:
\begin{equation}
\chi_{m,l}=\frac{\partial^2 \tilde f}{\partial m_l^2},~~
\chi_{t,l}=\frac{\partial^2 \tilde f}{\partial t \partial m_l },~~
\chi_{t,t}=\frac{\partial^2 \tilde f}{\partial t^2}.
\end{equation}
Thus three different  pseudo-critical temperatures $T_{m,l}$, $T_{t,l}$ and $T_{t,t}$ can be defined. 
In the vicinity of
the critical point the behavior of these susceptibilities is controlled by three universal
scaling function that can be derived from $f_f$. In the chiral limit $T_{m,l}=T_{t,l}=T_{t,t}=T_c^0$. 
There is, however, an additional complication for $O(N)$ universality class:  while $\chi_{m,l}$ and $\chi_{t,l}$
diverge at the critical point for $m_l \rightarrow 0$
\begin{equation}
\chi_{m,l} \sim m_l^{1/\delta - 1},~~~
\chi_{t,l} \sim m_l^{(\beta -1)/\beta\delta}, 
\label{peaks}
\end{equation}
$\chi_{t,t}$ is finite because $\alpha<0$ for $O(N)$ models ( $\chi_{t,t} \sim |t|^{-\alpha}$  ).
Therefore, one has to consider the third derivative of $\tilde f$ with respect to $t$ :
\begin{equation}
\chi_{t,t,t}=\frac{\partial^3 \tilde f}{\partial t^3}.
\end{equation}

In the vicinity of the critical point the derivatives with respect to $t$ can be estimated 
by taking the derivatives with respect to $\mu_l^2$, i.e.
the response functions $\chi_{t,l}$ and $\chi_{t,t,t}$ are identical to the second Taylor expansion coefficient
of the quark condensate and the sixth order expansion coefficient to the pressure, respectively. The former
controls the curvature of the transition temperature as function of the quark chemical potential $\mu_q$ and
was studied for $p4$ action using $N_{\tau}=4$ and $8$ lattices \cite{curvature}. The later corresponds to the sixth
order quark number fluctuation which is related to the deconfinement aspects of the transition.
The fact that this quantity is sensitive to the chiral dynamics points to a relation between  deconfining and chiral
aspects of the transition. 
In the following I discuss the determination of the transition temperature defined as
peak position of $\chi_{m,l}$, i.e. $T_c=T_{m,l}$. 

\subsection{Determination of the transition temperature}
The $O(N)$ scaling described in the above subsection can be used to determine the pseudo-critical temperature of
the chiral transition. For the study of the $O(N)$ scaling it is convenient to consider the dimensionless 
order parameter
\begin{equation}
M_b=m_s \frac{\langle \bar \psi \psi\rangle_l}{T^4}.
\end{equation}
The subscript "b" refers to the fact that this is a bare quantity since the additive UV divergence is not removed.
From the point of view of the scaling analysis this divergent term is just a regular contribution.
For sufficiently small quark mass and in the vicinity of the transition region we can write 
\begin{equation}
M_b(T,H) = h^{1/\delta} f_G(t/h^{1/\beta\delta}) + f_{M,reg}(T,H). 
\label{order_scaling}
\end{equation}
Here $f_G(z)$ is the scaling function related to $f_f$ and was calculated for $O(2)$ and $O(4)$ spin models
\cite{EngelsO2,Toussaint,EngelsO4}.
The regular contribution can be parametrized as \cite{hotqcd2}
\begin{eqnarray}
f_{M,reg}(T,H) &=& a_t(T)  H   \nonumber \\
&=& \left( a_0 + a_1 \frac{T-T_c^0}{T_c^0} + a_2 \left(\frac{T-T_c^0}{T_c^0} \right)^2 \right) H.
\label{eq:freg}
\end{eqnarray}
Then we have the following behavior for the light chiral susceptibility
\begin{eqnarray}
\frac{\chi_{m,l}}{T^2} &=& \frac{T^2}{m_s^2}
\left( \frac{1}{h_0}
h^{1/\delta -1} f_\chi(z) + \frac{\partial f_{M,reg}(T,H)}{\partial H}
\right) \; , \nonumber \\
&&{\rm with}\;\; f_{\chi}(z)=\frac{1}{\delta} [f_G(z)-\frac{z}{\beta} f_G'(z)].
\label{eq:chiralsuscept}
\end{eqnarray}
One then performs a simultaneous fit to the lattice data for $M_b$ and $\chi_{m,l}$ treating
$T_c^0, t_0, h_0, a_0, a_1$ and $a_2$ as fit parameters \cite{hotqcd2}. This gives a good description
of the quark mass and temperature dependence of $\chi_{m,l}$ and allows to determine
accurately the peak position in $\chi_{m,l}$.
Using this scaling analysis $T_c$ has been determined for $asqtad$ and $HISQ$ actions for different $N_{\tau}$.
Having determined $T_c$ for 
$HISQ$  and $asqtad$ action for each $N_{\tau}$ a combined continuum extrapolation was performed using different
assumption about the $N_{\tau}$ dependence of $T_c$ which resulted in the value \cite{hotqcd2}:
\begin{equation}
T_c=(159 \pm 9)~ {\rm MeV}.
\label{tc}
\end{equation}
The analysis also demonstrated that $HISQ$ and $asqtad$ action give consistent results in the continuum limit.
The Budapest-Wuppertal collaboration found $T_c=147(2)(3)$MeV, $157(3)(3)$MeV and 
$155(3)(3)$MeV defined as peak position in the chiral susceptibility, inflection points in $\Delta_{l,s}$ 
and $\Delta_l^R$ respectively \cite{fodor10}. These agree with the above value within errors. The peak position
in $\chi_{disc}$ calculated using Domain Wall Fermions is also consistent with the $T_c$ value in Eq. (\ref{tc}).

\section{Conclusions}

In recent years significant progress has been made in lattice QCD calculations at non-zero temperature. Chiral
and deconfining aspects of the QCD transition have been studied using improved staggered quark formulation
allowing to control discretization effects. Some quantities have been calculated at small baryon density 
using Taylor expansion in chemical potentials. 
At sufficiently low temperatures lattice results can be understood in terms of hadron resonance gas model,
while at high temperatures resummed perturbative calculations describe the lattice data quite well. 
For several quantities it has been shown that in the continuum limit different discretization
schemes, including discretizations other than staggered, give consistent results. 
In particular, agreement has been reached on the value of the chiral transition
temperature. There is still disagreement in the lattice calculation of the equation of state.

\end{document}